\documentclass{iopart}

\usepackage{graphics}
\usepackage{epsf}
\usepackage{epsfig}

\begin{document} 


\title[Crossover scaling in the Domany-Kinzel cellular automaton]{Crossover 
scaling in the Domany-Kinzel cellular automaton}

\author{S. L\"ubeck}

\address{
Institut f\"ur Theoretische Physik (B), 
RWTH Aachen,\\
52056 Aachen, Germany 
}

\ead{sven.luebeck@physik.rwth-aachen.de}

{\vspace{-6.0cm}
\noindent
accepted for publication in 
{\it Journal of Statistical Mechanics, (2006)}
\vspace{5.5cm}}


\begin{abstract}
We consider numerically the crossover scaling behavior from the
directed percolation universality class to the 
compact directed percolation universality class 
within the one-dimensional Domany-Kinzel cellular automaton.
Our results are compared to those of a recently performed 
field theoretical approach.
In particular, the value of the crossover exponent $\phi=2$
is confirmed.
\end{abstract}



\section{Introduction}

We consider the crossover scaling behavior between two
non-equilibrium universality classes.
Both universality classes describe second order phase
transitions into an absorbing state where the dynamics
of the model is trapped for ever.
The first class is the famous directed percolation (DP)
universality class 
(see \cite{MARRO_1,HINRICHSEN_1,ODOR_1,LUEB_35}
for recent reviews). 
The second class is the so-called compact directed 
percolation (CDP) universality class which is also known as the 
voter universality class within the mathematical literature.
Whereas the one-dimensional CDP behavior is well
understood due to a mapping to random walks, 
the stochastic process of DP is still analytically unsolved.
Numerical investigations of this crossover trace back 
to~\cite{MENDES_2}.
Here, we numerically examine 
the crossover between both universality classes 
within the one-dimensional Domany-Kinzel (DK) cellular 
automaton~\cite{DOMANY_1}.
This model is well suited for the corresponding crossover 
investigation since its phase diagram exhibits a line of
DP-like transitions, terminating in a CDP endpoint.
In particular, we consider the temporal evolution of the
order parameter.
The behavior of this quantity reflects the crossover and will
be useful to determine the crossover exponent.
Our results confirm those of a recently performed field theoretical
approach~\cite{JANSSEN_4}.
In particular, the value of the crossover exponent, which is 
expected to be exact, is confirmed.

\section{Directed Percolation}
\label{sec:directed_perc}

The stochastic process of directed percolation (DP) can be 
considered as a paradigm of non-equilibrium phase transitions 
into an absorbing state (see \cite{MARRO_1,HINRICHSEN_1,ODOR_1,LUEB_35}
for recent reviews).
The process of directed percolation can be 
represented by the Langevin equation~\cite{JANSSEN_1} 
(in the Ito sense, see e.g.~\cite{KAMPEN_1}) 
\begin{equation}
\lambda^{-1} \,
\partial_{\scriptscriptstyle t} \, \rho_{\scriptscriptstyle\mathrm{a}}({\mathbf{r}},t)
\; = \; \tau \, \rho_{\scriptscriptstyle\mathrm{a}}({\mathbf{r}},t) 
\, - \, u \, \rho_{\scriptscriptstyle\mathrm{a}}({\mathbf{r}},t)^2 
\, + \, \nabla^2 \, \rho_{\scriptscriptstyle\mathrm{a}}({\mathbf{r}},t) 
\, + \, 
\eta({\mathbf{r}},t) 
\label{eq:langevin_dp_01}
\end{equation}
which describes the order parameter 
$\rho_{\scriptscriptstyle\mathrm{a}}({\mathbf{r}},t)$,
i.e., the density of active sites, on a mesoscopic scale.
A positive order parameter occurs above the
transition point ($\tau=0$) whereas the absorbing state 
$\rho_{\scriptscriptstyle\mathrm{a}}=0$
is approached for negative $\tau$, below the transition point.
Furthermore, $\eta$ denotes the 
noise which accounts for fluctuations of the particle
density~$\rho_{\scriptscriptstyle\mathrm{a}}({\mathbf{r}},t)$.
According to the central limit theorem, 
$\eta({\mathbf{r}},t)$ 
is a Gaussian random variable with zero mean
and whose correlator is given by
\begin{equation}
\langle \, \eta({\mathbf{r}},t) \, 
\eta({\mathbf{r}}^{\prime},t^{\prime}) \, \rangle
\; = \; \lambda^{-1} \,
\Gamma \; \rho_{\scriptscriptstyle\mathrm{a}}({\mathbf{r}},t) \;
\delta({\mathbf{r}}-{\mathbf{r}}^{\prime}) \;
\delta(t-t^{\prime}) \, .
\label{eq:langevin_dp_corr_01}
\end{equation}
Notice that the noise ensures that the 
system is trapped in the absorbing state 
\mbox{$\rho_{\scriptscriptstyle\mathrm{a}}({\mathbf{r}},t)=0$}.
Furthermore, higher order terms such as 
$\rho_{\scriptscriptstyle\mathrm{a}}({\mathbf{r}},t)^3,
\rho_{\scriptscriptstyle\mathrm{a}}({\mathbf{r}},t)^4, \ldots$ 
(or 
$\nabla^4\rho_{\scriptscriptstyle\mathrm{a}}
({\mathbf{r}},t),
\nabla^6\rho_{\scriptscriptstyle\mathrm{a}}({\mathbf{r}},t), \ldots$)
are irrelevant under renormalization group transformations
as long as $u > 0$.
Negative values of $u$ give rise to a first order
phase transition whereas $u=0$ is associated with
a tricritical\index{tricritical point} point 
(see \cite{OHTSUKI_1,OHTSUKI_2,JANSSEN_4} for field theoretical as well
as \cite{LUEB_31,GRASSBERGER_13} for recently performed 
numerical investigations of tricritical directed percolation).

As usual for second order phase transitions the system obeys certain scaling 
laws close to criticality (see \cite{HINRICHSEN_1} or \cite{LUEB_35}
for complete discussions of the critical scaling behavior
of absorbing phase transitions).
For example, the steady state order parameter scales (up to higher orders)
as 
\begin{equation}
\rho_{\scriptscriptstyle\mathrm{a}} \, \propto \, \tau^{\beta} \, 
\label{eq:ord_par_scal}
\end{equation}
for $\tau >0$.
Additionally to the steady state scaling behavior, the dynamical
scaling is also expressed in terms of power laws
\begin{equation}
\rho_{\scriptscriptstyle\mathrm{a}}(t) \, \propto \, t^{-\alpha} \, ,
\qquad 
P_{\scriptscriptstyle\mathrm{a}}(t) \, \propto \, t^{-\delta} \, .
\label{eq:dyna_scal}
\end{equation}
The first law describes the order parameter decay at criticality 
starting from a fully occupied lattice.
The second power law describes the temporal evolution of the 
survival probability~$P_{\scriptscriptstyle\mathrm{a}}$ of an 
initially isolated single seed of activity.  
This probability is related to the probability
that a given site belongs to a percolating cluster 
\begin{equation}
P_{\scriptscriptstyle\mathrm{perc}} \, \propto \, \tau^{\beta^{\prime}} \, ,
\label{eq:perc_prob}
\end{equation}
for $\tau>0$.
Often this quantity is considered as the order parameter of the
percolation transition, additionally to the steady state 
density~$\rho_{\scriptscriptstyle\mathrm{a}}$.
Above the upper critical dimension $d_{\scriptscriptstyle\mathrm{c,DP}}=4$, 
the exponents equal their
mean field values, e.g.~$\beta=1$ and $\beta^{\prime}=1$.

Below the upper critical dimension 
renormalization group techniques have to be applied
to determine the critical exponents.
In that case path integral formulations\index{path integral} are more
adequate than the Langevin equation approach~(see e.g.~\cite{CHAICHIAN_1}).
Furthermore, they provide a deeper understanding of the 
stochastic process.
Stationary correlation functions as well as response
functions can be determined by calculating 
path integrals with weight $\exp{(-{\mathcal J})}$,
where the dynamic functional ${\mathcal J}$ describes
the considered stochastic process.
Up to higher irrelevant orders the dynamic functional 
associated with directed percolation is given 
by~\cite{JANSSEN_1,JANSSEN_7,DEDOMINICIS_1,JANSSEN_8} 
\begin{equation}
{\mathcal J}
[{\tilde \rho_{\scriptscriptstyle\mathrm{a}}},\rho_{\scriptscriptstyle\mathrm{a}}] 
=  \lambda \int
{\mathrm d}^d{\mathbf{r}} \,
{\mathrm d}t \,
{\tilde \rho_{\scriptscriptstyle\mathrm{a}}}  \left [ \lambda^{-1}
\partial_{\scriptscriptstyle t} \rho_{\scriptscriptstyle\mathrm{a}}  
-  (\tau + \nabla^2 ) \rho_{\scriptscriptstyle\mathrm{a}}
 -  \left (  \frac{\Gamma}{2}  {\tilde \rho_{\scriptscriptstyle\mathrm{a}}}  
- u \rho_{\scriptscriptstyle\mathrm{a}} \right ) 
\rho_{\scriptscriptstyle\mathrm{a}} 
\right ]
\label{eq:action_reggeon_field_theory}
\end{equation}
where ${\tilde \rho_{\scriptscriptstyle\mathrm{a}}}({\mathbf{r}},t)$
denotes the response field conjugated to the 
Langevin noise field~\cite{MARTIN_1}.
The functional $J$ is invariant under the duality
transformation (so-called rapidity reversal in Reggeon field 
theory)
\begin{equation}
\mu^{-1} \,
{\tilde \rho_{\scriptscriptstyle\mathrm{a}}}({\mathbf{r}},t) 
\longleftrightarrow
\, - \mu \, {\rho_{\scriptscriptstyle\mathrm{a}}}({\mathbf{r}},-t) 
\label{eq:rapidity_trans}
\end{equation}
with the redundant variable $\mu^2=2u/ \Gamma$.
As usual, the duality transformation defines a 
dual stochastic process that might differ from the original one~\cite{LIGGETT_1}.
At criticality, the average density of particles 
of the dual process 
$\rho_{\scriptscriptstyle\mathrm{a}}^{\scriptscriptstyle \mathrm{dual}}$ 
equals asymptotically the survival probability~\cite{JANSSEN_4} via
\begin{equation}
P_{\scriptscriptstyle{\mathrm{a}}}(t) \; \sim \; \mu^2 \, 
\rho_{\scriptscriptstyle\mathrm{a}}^{\scriptscriptstyle \mathrm{dual}}(t) \, .
\label{eq:p_sur_alpha_rho_dual}
\end{equation}
Here, $P_{\scriptscriptstyle{\mathrm{a}}}(t)$ denotes again the probability 
that a cluster generated by a single seed is still active
after $t$ time steps.
On the other hand, $\rho_{\scriptscriptstyle\mathrm{a}}^{\scriptscriptstyle \mathrm{dual}}(t)$
describes the particle decay of the dual process started from
a fully occupied lattice.
The self-duality of directed percolation 
implies \mbox{$\rho_{\scriptscriptstyle\mathrm{a}}(t)
=\rho_{\scriptscriptstyle\mathrm{a}}^{\scriptscriptstyle \mathrm{dual}}(t)$}
and therefore~\cite{GRASSBERGER_4,JANSSEN_4}  
\begin{equation}
P_{\scriptscriptstyle{\mathrm{a}}}(t) \; 
\sim \; \mu^2 \, \rho_{\scriptscriptstyle\mathrm{a}}(t) \, .
\label{eq:p_sur_alpha_rho}
\end{equation}
The asymptotic equivalence ensures that both quantities 
have the same exponents~\cite{GRASSBERGER_4},
including
\begin{equation}
\alpha \; = \; \delta \, . 
\label{eq:alpha_delta_equal}
\end{equation}
Taking the scaling laws $\beta=\nu_{\scriptscriptstyle \parallel} \alpha$
and $\beta^{\prime}=\nu_{\scriptscriptstyle \parallel} \delta $
into account ($\nu_{\scriptscriptstyle \parallel}$ denotes the exponent of
the temporal correlations) equation\,(\ref{eq:alpha_delta_equal})
implies the identity 
\begin{equation}
\beta \; = \; \beta^{\prime} 
\label{eq:beta_betap_equal}
\end{equation}
for the universality class of directed percolation.

It is worth mentioning that equation\,(\ref{eq:p_sur_alpha_rho})
follows from the rapidity reversal symmetry of the dynamical
functional~${\mathcal{J}}$, i.e., it is a specific
property of the directed percolation universality class.
Thus, compared to general absorbing phase transitions,
the number of independent critical exponents
for directed percolation is reduced.
Furthermore, the self-duality is expressed within the field theoretical
treatment of the minimal Langevin equation of directed percolation.
Hence, the important rapidity reversal
is usually reflected on a coarse grained level only
and holds often only asymptotically.
In other words, it is often masked on a microscopic level, i.e.~it
does not necessarily represent a physical symmetry 
of microscopic models~\cite{JANSSEN_1}.
Nevertheless, the robustness and the ubiquity 
of the DP universality class is
expressed by the conjecture of Janssen and 
Grassberger~\cite{JANSSEN_1,GRASSBERGER_2}:
short-range interacting systems, exhibiting a
continuous phase transition into an absorbing state,
belong to the DP universality class, if they are characterized
by a one-component order parameter.
Different universal scaling behavior is expected to occur
in the presence of additional symmetries or relevant disorder effects.

\section{Domany-Kinzel automaton and Compact Directed Percolation}
\label{sec:dk_auto}

\begin{figure}[t] 
\centering
\includegraphics[clip,width=11.0cm,angle=0]{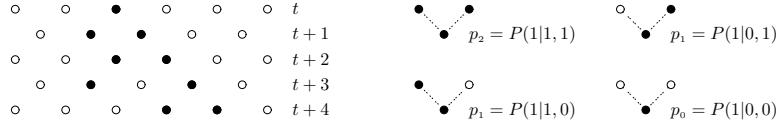}
\caption[Domany-Kinzel automaton] 
{The $1+1$-dimensional Domany-Kinzel automaton.
Occupied sites are marked by full circles.
A spreading of particles is sketched in the left part.
The state of the one-dimensional system at time $t+1$ is
obtained by an iteration of the dynamics rules
that are embodied in the conditional probabilities
$P(n_{\scriptscriptstyle t+1}|n_{{\scriptscriptstyle \mathrm {l},t}},
n_{{\scriptscriptstyle \mathrm {r},t}})$.
Thus the occupation $n_{\scriptscriptstyle t+1}$ of a given site at time~$t$
depends on the state of the left (l) and right (r) backward neighbors 
($n_{{\scriptscriptstyle \mathrm {l},t}},n_{{\scriptscriptstyle \mathrm {r},t}}$) 
at time $t$.
Spontaneous particle creation corresponds to the conjugated field
and is usually forbidden, i.e., $p_{\scriptscriptstyle 0}=0$.
\label{fig:dk_model_01}} 
\end{figure}

Several lattice models are known which exhibit an absorbing phase
transition belonging to the directed percolation universality 
class.
Famous examples are the contact process~\cite{HARRIS_2}, 
the pair-contact process~\cite{JENSEN_2},
the Ziff-Gulari-Barshad model~\cite{ZIFF_1}, 
and the threshold transfer process~\cite{MENDES_1}
(see e.g.~\cite{LUEB_32} for a detailed scaling analysis of various
DP-like lattice models).
Another well known $1+1$-dimensional stochastic cellular automaton 
exhibiting directed percolation scaling behavior is 
the Domany-Kinzel (DK) automaton~\cite{DOMANY_1}.
It is defined on a diagonal square lattice where the lattice sites
are either empty ($n=0$) or occupied by a particle ($n=1$).
At discrete time steps a parallel update procedure is performed 
according to the following rules (see figure~\ref{fig:dk_model_01}).
A site at time $t$ is occupied with 
probability~$p_{\scriptscriptstyle 2}$ 
($p_{\scriptscriptstyle 1}$) if both (only one) backward
sites (at time $t-1$) are occupied.
Otherwise the site remains empty.

\begin{figure}[t]
  \centering
  \includegraphics[clip,width=7.0cm,angle=0]{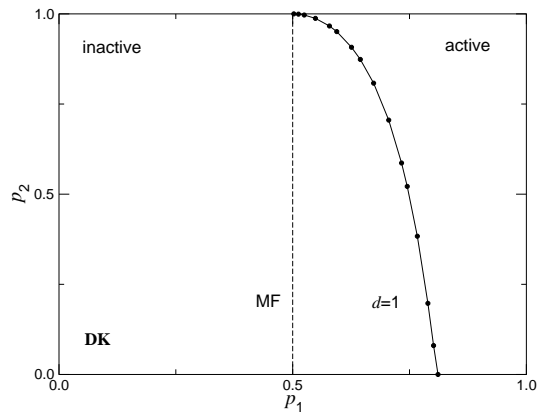}
  \caption{
     The phase diagram of the $1+1$-dimensional Domany-Kinzel automaton.
     The bold circles indicate the numerically determined transition
     points (the solid line is just to guide the eye). 
     The dashed line corresponds to the simplest mean field
     approximation (see text).
     In both cases, all transition points along the critical lines 
     belong to the
     universality class of directed percolation except of the
     termination point 
     ($p_{\scriptscriptstyle 1}=1/2$, $p_{\scriptscriptstyle 2}=1$).    
   }
  \label{fig:dk_phase_dia_01} 
\end{figure}

Notice that the Domany-Kinzel automaton depends on the two 
parameters $p_{\scriptscriptstyle 1}$ and $p_{\scriptscriptstyle 2}$.
This dependence results in a non-trivial phase diagram which is
shown in figure\,\ref{fig:dk_phase_dia_01}.
Some particular processes are included in the Domany-Kinzel
automaton.
For example, the process of bond directed percolation corresponds
to the choice $p_{\scriptscriptstyle 2}=p_{\scriptscriptstyle 1}
(2-p_{\scriptscriptstyle 1})$
whereas site directed percolation is obtained for
$p_{\scriptscriptstyle 1}=p_{\scriptscriptstyle 2}$~\cite{DOMANY_1}.
Furthermore, for $p_{\scriptscriptstyle 2}=0$ the Domany-Kinzel 
automaton equals to the cellular automata rule "18" of Wolfram's 
classification scheme~\protect\cite{WOLFRAM_1,ZEBENDE_1}.

Instructive insights into the behavior of the system can be 
gained already from a simple mean field 
treatment (see e.g.~\cite{KINZEL_2,TOME_1,RIEGER_1,ATMAN_1}).
Within a one-site approximation the above presented 
reaction scheme leads to the differential equation for the 
density of active (occupied) sites (see~\cite{LUEB_35} for
a detailed consideration)
\begin{equation}
\partial_{\scriptscriptstyle t} \, 
\rho_{\scriptscriptstyle\mathrm{a}}(p_{\scriptscriptstyle 1}, p_{\scriptscriptstyle 2}) 
\; = \; 2 \, (2 p_{\scriptscriptstyle 1}-1) \, \rho_{\scriptscriptstyle\mathrm{a}} 
\, - \, 2 \, (2 p_{\scriptscriptstyle 1} - p_{\scriptscriptstyle 2}) \,
\rho_{\scriptscriptstyle\mathrm{a}}^2 \, .
\label{eq:DK_mean_Langevin}
\end{equation}
Focusing on the steady state behavior 
($\partial_{\scriptscriptstyle t}\rho_{\scriptscriptstyle\mathrm{a}}=0$) 
we find that the order parameter is given by
\begin{equation}
\rho_{\scriptscriptstyle\mathrm{a}}(p_{\scriptscriptstyle 1}, 
p_{\scriptscriptstyle 2})\; = \; 0
 \quad\quad \vee
\quad\quad
\rho_{\scriptscriptstyle\mathrm{a}}
(p_{\scriptscriptstyle 1}, p_{\scriptscriptstyle 2}) 
\; = \; \frac{2 p_{\scriptscriptstyle 1}-1}{\,2 p_{\scriptscriptstyle 1} 
- p_{\scriptscriptstyle 2}\,} \, .
\label{eq:DK_mean_field_op_h0}
\end{equation}
The first solution is stable for $p_{\scriptscriptstyle 1}<1/2$ and
corresponds to the inactive phase.
The active phase is described by the second solution 
which is stable for $p_{\scriptscriptstyle 1}>1/2$.
The phase diagram of the 
Domany-Kinzel automaton exhibits two phases separated by the
critical line 
$p_{\scriptscriptstyle 1,\mathrm{c}}=1/2$
(see figure\,\ref{fig:dk_phase_dia_01}).
Along this critical line the order parameter vanishes
linearly ($\beta_{\scriptscriptstyle\mathrm{DP}}=1$) in the active phase 
\begin{equation}
\rho_{\scriptscriptstyle\mathrm{a}}(p_{\scriptscriptstyle 1}, p_{\scriptscriptstyle 2})
\; = \; \frac{1}{1-p_{\scriptscriptstyle 2}} \, \delta p
\, + \, {\mathcal O}(\delta p^2)
\label{eq:DK_mean_field_op_h0_scal}
\end{equation}
with $\delta p=(p_{\scriptscriptstyle 1} - p_{\scriptscriptstyle\mathrm{c}})
/p_{\scriptscriptstyle\mathrm{c}}$.
Thus, the critical exponent of the order parameter equals the mean field 
value of directed percolation as along as $p_{\scriptscriptstyle 2}<1$,
i.e., all critical points on the transition line (except of the termination
point $p_{\scriptscriptstyle 2}=1$) belong to the universality
class of directed percolation.
The last statement is also valid for the $1+1$-dimensional Domany-Kinzel
automaton although the transition line as well as the values of the 
critical exponents differ from the mean field 
approximations (see figure\,\ref{fig:dk_phase_dia_01}).

Clearly, the above derived mean field results are not valid
for $p_{\scriptscriptstyle 2}=1$. 
In that case no particle annihilation takes place
within a domain of occupied sites.
Thus, creation and annihilation processes are bounded to the domain 
walls where empty and occupied sites are adjacent.
This corresponds to the particle-hole symmetry
\begin{equation}
n \; \; \leftrightarrow \; \; 1-n
\label{eq:particle_hole_symmetry}
\end{equation}
which is also reflected in the mean field differential
equation\,(\ref{eq:DK_mean_Langevin})
\begin{equation}
\partial_{\scriptscriptstyle t} \, 
\rho_{\scriptscriptstyle\mathrm{a}}(p_{\scriptscriptstyle 1}, 
p_{\scriptscriptstyle 2}=1)  
\, = \, 2 \, (2 p_{\scriptscriptstyle 1}-1) \, \rho_{\scriptscriptstyle\mathrm{a}}
\, (1-\rho_{\scriptscriptstyle\mathrm{a}}) 
\, = \, 2 \, \delta p \, \rho_{\scriptscriptstyle\mathrm{a}}
\, (1-\rho_{\scriptscriptstyle\mathrm{a}}) \, . 
\label{eq:DK_mean_Langevin_CDP}
\end{equation}
To be precise, equation\,(\ref{eq:DK_mean_Langevin_CDP}) is
invariant under the transformation 
$\rho_{\scriptscriptstyle\mathrm{a}}\leftrightarrow
1-\rho_{\scriptscriptstyle\mathrm{a}}$ and 
$\delta p \leftrightarrow -\delta p$.
Beside the empty lattice, 
the fully occupied lattice is now another absorbing state.
The behavior of the domain wall boundaries can be mapped to 
a simple random walk~\cite{DOMANY_1} and the domains
of particles
grow on average for $\delta p> 0$ whereas they
shrink for $\delta p < 0$.
Thus the steady state density $\rho_{\scriptscriptstyle\mathrm{a}}$ is zero 
below \mbox{$p_{\scriptscriptstyle \mathrm c}=1/2$} 
and \mbox{$\rho_{\scriptscriptstyle\mathrm{a}}=1$} above
\mbox{$p_{\scriptscriptstyle \mathrm c}$}.
At the critical value the order parameter 
$\rho_{\scriptscriptstyle\mathrm{a}}$ exhibits a jump. 
The associated critical 
exponent $\beta_{\scriptscriptstyle \mathrm {CDP}}=0$
differs in all dimensions from the directed percolation 
values $\beta_{\scriptscriptstyle \mathrm {DP}}$.
Since domain branching does not take place  
the dynamical process for $p_{\scriptscriptstyle 2}=1$ is often
termed compact directed percolation (CDP).
The critical behavior equals that of the $1+1$-dimensional 
voter model~\cite{LIGGETT_1} and it is analytically tractable 
due to the mapping to random walks.
Exact results are derived for the critical exponents~\cite{DOMANY_1,ESSAM_1} 
($\beta=0$, $\beta^{\prime}=1$, $\nu_{\parallel}=2$, $\nu_{\perp}=1$), 
as well as for certain finite-size scaling functions~\cite{KEARNEY_1}.
In particular, the domain growth from a single seed exhibits a 
one-to-one correspondence to a pair of annihilating random walkers.
That correspondence allows the calculation of the percolation 
probability~\cite{ESSAM_1} 
\begin{equation}
P_{\scriptscriptstyle\mathrm{perc}}(p_{\scriptscriptstyle 1}) \; = \;
\left \{
\begin{array}{lcl}
0 & \mathrm{if} & p_{\scriptscriptstyle 1}<1/2 \, ,\\[2mm]
\frac{2\, p_{\scriptscriptstyle 1}-1}{p_{\scriptscriptstyle 1}^2} 
& \mathrm{if} & p_{\scriptscriptstyle 1}>1/2 \, ,
\end{array}
\right .
\label{eq:CDP_Pperc}
\end{equation}
yielding the percolation 
exponent $\beta^{\prime}_{\scriptscriptstyle\mathrm{CDP}}=1$.
In contrast to directed percolation the 
universality class of compact directed 
percolation\index{universality class, CDP} 
is\index{universality class, compact directed percolation} 
characterized by the inequality
\begin{equation}
\beta_{\scriptscriptstyle\mathrm{CDP}} \; 
\neq \; \beta^{\prime}_{\scriptscriptstyle\mathrm{CDP}} \, .
\label{eq:beta_betap_nequal}   
\end{equation}
The number of independent critical exponents is therefore
greater than for directed percolation. 
In summary, the universality class of compact directed
percolation is characterized by a continuously vanishing
percolation order parameter $P_{\scriptscriptstyle\mathrm{perc}}$ and algebraically
diverging correlations lengths~\cite{ESSAM_1},
indicating a second order phase transition.
But due to the misleading discontinuous behavior 
of the steady state order parameter~$\rho_{\scriptscriptstyle\mathrm{a}}$ 
the phase transition was often considered as first order.

Within a field theoretical approach, the CDP process can be
described by the Langevin equation~\cite{JANSSEN_4}
\begin{equation}
\label{eq:langevin_cdp}   
\lambda^{-1}\, \partial_{\scriptscriptstyle t} 
\rho_{\scriptscriptstyle\mathrm{a}}
= \tau  \rho_{\scriptscriptstyle\mathrm{a}} 
(1-\rho_{\scriptscriptstyle\mathrm{a}})
+ \nabla^2  \rho_{\scriptscriptstyle\mathrm{a}}
+ \eta  \, .
\label{eq:langevin_CDP} 
\end{equation}
In order to ensure that the fully occupied and empty lattice is
absorbing the noise correlator obeys
\begin{equation}
\label{eq:langevin_cdp_corr}   
\langle \eta({\mathbf{r}},t)  
\eta({\mathbf{r}}^{\prime},t^{\prime})  \rangle
=  \lambda^{-1} \Gamma  \, \rho_{\scriptscriptstyle\mathrm{a}}({\mathbf{r}},t) \, 
[1-\rho_{\scriptscriptstyle\mathrm{a}}({\mathbf{r}},t)] \, 
\, \delta({\mathbf{r}}-{\mathbf{r}}^{\prime}) 
\, \delta(t-t^{\prime})  .
\end{equation}
Simple dimensional counting shows that the noise is irrelevant
for $d>2$, i.e., the value of the
upper critical dimension is $d_{\scriptscriptstyle\mathrm{c,CDP}}=2$.
Associated to this Langevin equation is the response functional~\cite{JANSSEN_4}
\begin{equation}
{\mathcal J}[{\tilde \rho_{\scriptscriptstyle\mathrm{a}}},
\rho_{\scriptscriptstyle\mathrm{a}}]  =  \lambda  \int
{\mathrm d}^{\scriptscriptstyle d}{\mathbf{r}} \,  
{\mathrm d}t \,
{\tilde \rho_{\scriptscriptstyle\mathrm{a}}}   \left [  
\lambda^{-1}\partial_{\scriptscriptstyle t} \rho_{\scriptscriptstyle\mathrm{a}}  
-  \tau \rho_{\scriptscriptstyle\mathrm{a}}(1-\rho_{\scriptscriptstyle\mathrm{a}})
- \nabla^2  \rho_{\scriptscriptstyle\mathrm{a}}
-  \frac{\Gamma}{2}  {\tilde \rho_{\scriptscriptstyle\mathrm{a}}}  
\rho_{\scriptscriptstyle\mathrm{a}} (1-\rho_{\scriptscriptstyle\mathrm{a}}) 
\right ] .
\label{eq:action_comp_dp}
\end{equation}
The functional is invariant under the transformation
\begin{eqnarray}
\label{eq:cdp_trans}
{\rho_{\scriptscriptstyle\mathrm{a}}}(\mathbf{r},t) & \longleftrightarrow &
\, 1\, -\, {\rho_{\scriptscriptstyle\mathrm{a}}}(\mathbf{r},-t) \, , \nonumber \\[0mm]
{\tilde \rho_{\scriptscriptstyle\mathrm{a}}}(\mathbf{r},t) & \longleftrightarrow &
\, -\, {\tilde \rho_{\scriptscriptstyle\mathrm{a}}}(\mathbf{r},-t) \, , \\[0mm]
{\tau} & \longleftrightarrow &
-\tau \, , \nonumber 
\end{eqnarray}
which reflects the particle-hole symmetry and differs obviously 
from the rapidity-reversal symmetry of directed percolation.

In the following we consider the crossover scaling from
CDP to DP.
It is instructive to express the mean field equation\,(\ref{eq:DK_mean_Langevin})
in terms of the distances to the critical line $\tau=2\delta p$
and to the termination point 
$\kappa=2(1-p_{\scriptscriptstyle 2})$
\begin{equation}
\partial_{\scriptscriptstyle t} \, 
\rho_{\scriptscriptstyle\mathrm{a}}(p_{\scriptscriptstyle 1}, p_{\scriptscriptstyle 2}) 
\, = \, \tau \, \rho_{\scriptscriptstyle\mathrm{a}} 
(1-\rho_{\scriptscriptstyle\mathrm{a}} )
\, - \, \kappa\, \rho_{\scriptscriptstyle\mathrm{a}}^2 \, .
\label{eq:DK_mean_Langevin_02}
\end{equation}
From the CDP point of view ($\kappa=0$),
the particle-hole symmetry is broken for any
positive $\kappa$, i.e., for any vertical distance from the
termination point in the phase diagram.
The crossover from the CDP to the DP behavior can be described
in terms of scaling forms.
For example, the steady state order parameter obeys for any 
positive value of $\lambda$
\begin{equation}
\rho_{\scriptscriptstyle\mathrm{a}}(\tau,\kappa) 
\, \sim \, \lambda^{-\beta_{\scriptscriptstyle \mathrm{CDP}}}
\, {\tilde r}(\tau \, \lambda, \kappa \, \lambda^{\phi})
\label{eq:scal_cross_01}
\end{equation}
with $\beta_{\scriptscriptstyle \mathrm{CDP}}=0$ 
and where $\phi$ denotes the crossover exponent.
Setting $\lambda = \kappa^{-1/\phi}$ we obtain
$\rho_{\scriptscriptstyle\mathrm{a}}(\tau,\kappa) 
= {\tilde r}(\tau \, \kappa^{-1/\phi},1)$.
Comparing this result with the steady state 
solution (see equation \ref{eq:DK_mean_field_op_h0})
\begin{equation}
\rho_{\scriptscriptstyle\mathrm{a}}(\tau,\kappa) \, = \,
\frac{\tau /\kappa}{1+\tau /\kappa}
\label{eq:steady_state_02}
\end{equation}
we can identify the crossover exponent as well as the
crossover scaling function
\begin{equation}
\phi_{\scriptscriptstyle\mathrm{MF}} \, = \, 1\, , 
\qquad {\tilde r}_{\scriptscriptstyle\mathrm{MF}}(x,1) \,= \, 
\frac{x}{1+x} \, .
\label{eq:cross_phi_scal_func}
\end{equation}

Studying the crossover beyond the mean field level,
we have to add to the Langevin equation\,(\ref{eq:langevin_CDP})
an appropriate term which breaks the particle hole-symmetry.
For example, within the above presented mean field 
approximation the particle-hole symmetry is broken by 
$-\kappa \rho_{\scriptscriptstyle\mathrm{a}}^2$ 
(see equation\,\ref{eq:DK_mean_Langevin_02}).
Alternatively, a term of linear 
order $\rho_{\scriptscriptstyle\mathrm{a}}$ acts in the same
way.
The latter case is examined in ref.\,\cite{JANSSEN_4} by Janssen.
Performing a field theoretical treatment, Janssen derived
the crossover exponent
\begin{equation}
\phi \, = \, \left \{
\begin{array}{lll} 
\frac{2}{d} & \mathrm{if} & d <   d_{\scriptscriptstyle\mathrm{c}}=2 \\[2mm]
1	    & \mathrm{if} & d \ge d_{\scriptscriptstyle\mathrm{c}}=2 \, .
\end{array}
\right .
\label{eq:crossover_exp_d}
\end{equation}
Remarkably, these field theoretical results are expected 
to be exact since the involved diagrams can be summed 
exactly (see \cite{JANSSEN_4} for a detailed discussion).
In the following we consider the $1+1$-dimensional Domany-Kinzel automaton
in the vicinity of the termination point and confirm numerically the
result $\phi_{\scriptscriptstyle d=1}=2$.

\section{Numerical Analysis of the Crossover Behavior}
\label{sec:num_analysis}

At the beginning of our analysis, we determine several
critical points in order to obtain an estimate of the transition line.
For this purpose, we consider the
dynamical properties of the system along various
lines $p_{\scriptscriptstyle 2}= k p_{\scriptscriptstyle 1}$
for $k$ ranging from $0.1$ up to $1.99$ (see table\,\ref{table:pc_werte}).
For each value, we vary $p_{\scriptscriptstyle 1}$ and determine
the critical point.
Therefor, we use a standard spreading 
procedure~\cite{GRASSBERGER_4}, i.e., we 
consider the spreading of an initial seed and measure the
survival probability $P_{\scriptscriptstyle \mathrm{a}}(t)$.
At the critical point $p_{\scriptscriptstyle 1,\mathrm{c}}$,
the survival probability obeys a power-law behavior with the critical
exponent $\delta=\alpha=0.159464(6)$\cite{JENSEN_5},
see equation\,(\ref{eq:dyna_scal}).
Slightly above or below the critical value, the survival
probability exhibits deviations from a pure power-law,
resulting in a significant (left or right) curvature in a 
double-logarithmic plot (not shown).
In this way we get a sufficient estimate of the transition line.
The corresponding data are shown in figure\,\ref{fig:dk_phase_dia_01}
and are listed in table\,\ref{table:pc_werte}.

\begin{figure}[t]
  \centering
  \includegraphics[clip,width=7.0cm,angle=0]{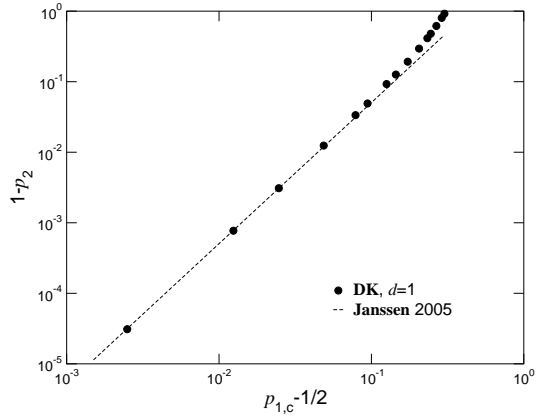}
  \caption{
     The transition points (bold circles) of the $1+1$-dimensional 
     Domany-Kinzel automaton 
     according to equation\,(\ref{eq:phase_line_cross_p}).
     The dashed line indicates the analytically expected
     behavior with the crossover exponent $\phi=2$.
   }
  \label{fig:dk_phase_dia_02} 
\end{figure}

As well known from scaling theory, the crossover exponent~$\phi$
determines, additionally to the crossover scaling behavior, the 
shape of the transition line~\cite{RIEDEL_2}.
%
%
In our case, the phase boundary is expected to obey asymptotically
\begin{equation}
\tau_{\scriptscriptstyle\mathrm{c}} \, \propto \, \kappa^{1/\phi} \, .
\label{eq:phase_line_cross}
\end{equation}
Here, the crossover parameter $\kappa$ equals again the distance to 
the termination point
along the $p_{\scriptscriptstyle 2}$-direction,
i.e., $\kappa \propto (1-p_{\scriptscriptstyle 2})$.
On the other hand, $\tau_{\scriptscriptstyle\mathrm{c}}$ is given by
\begin{equation}
\tau_{\scriptscriptstyle\mathrm{c}} \; \propto \; 
p_{\scriptscriptstyle 1,\mathrm{c}}(\kappa)-
p_{\scriptscriptstyle 1,\mathrm{c}}(\kappa=0)
\; = \; p_{\scriptscriptstyle 1,\mathrm{c}}(p_{\scriptscriptstyle 2})-\frac{1}{2} \, .
\label{eq:phase_line_cross_tau_def}
\end{equation}
Thus, the transition line behaves as
\begin{equation}
p_{\scriptscriptstyle 1,\mathrm{c}}(\kappa) \, = \, 
p_{\scriptscriptstyle 1,\mathrm{c}}(p_{\scriptscriptstyle 2}) \, = \, 
\frac{1}{2} \, + \, \mathrm{const} \, 
(1-p_{\scriptscriptstyle 2})^{1/\phi} \, .
\label{eq:phase_line_cross_p}
\end{equation}
Therefore, we plot in figure\,\ref{fig:dk_phase_dia_02} the crossover
parameter $1-p_{\scriptscriptstyle 2}$ as a function of 
$p_{\scriptscriptstyle 1,\mathrm{c}}-1/2$ in a double logarithmic plot.
Approaching the termination point, the transition line
agrees well with equation\,(\ref{eq:phase_line_cross})
and confirms the field theoretical result $\phi=2/d$~\cite{JANSSEN_4},
i.e., $\phi=2$ for $d=1$. 
But as can be seen, equation\,(\ref{eq:phase_line_cross})
describes only the asymptotic behavior close to the termination
point, i.e., for $\kappa \ll 0.1$.
This reflects the notorious difficulties to determine the
crossover exponent~$\phi$ from the phase boundary since it is
in general not clear whether the asymptotic scaling regime is 
actually reached.

\begin{figure}[t]
  \centering
  \includegraphics[clip,width=7.0cm,angle=0]{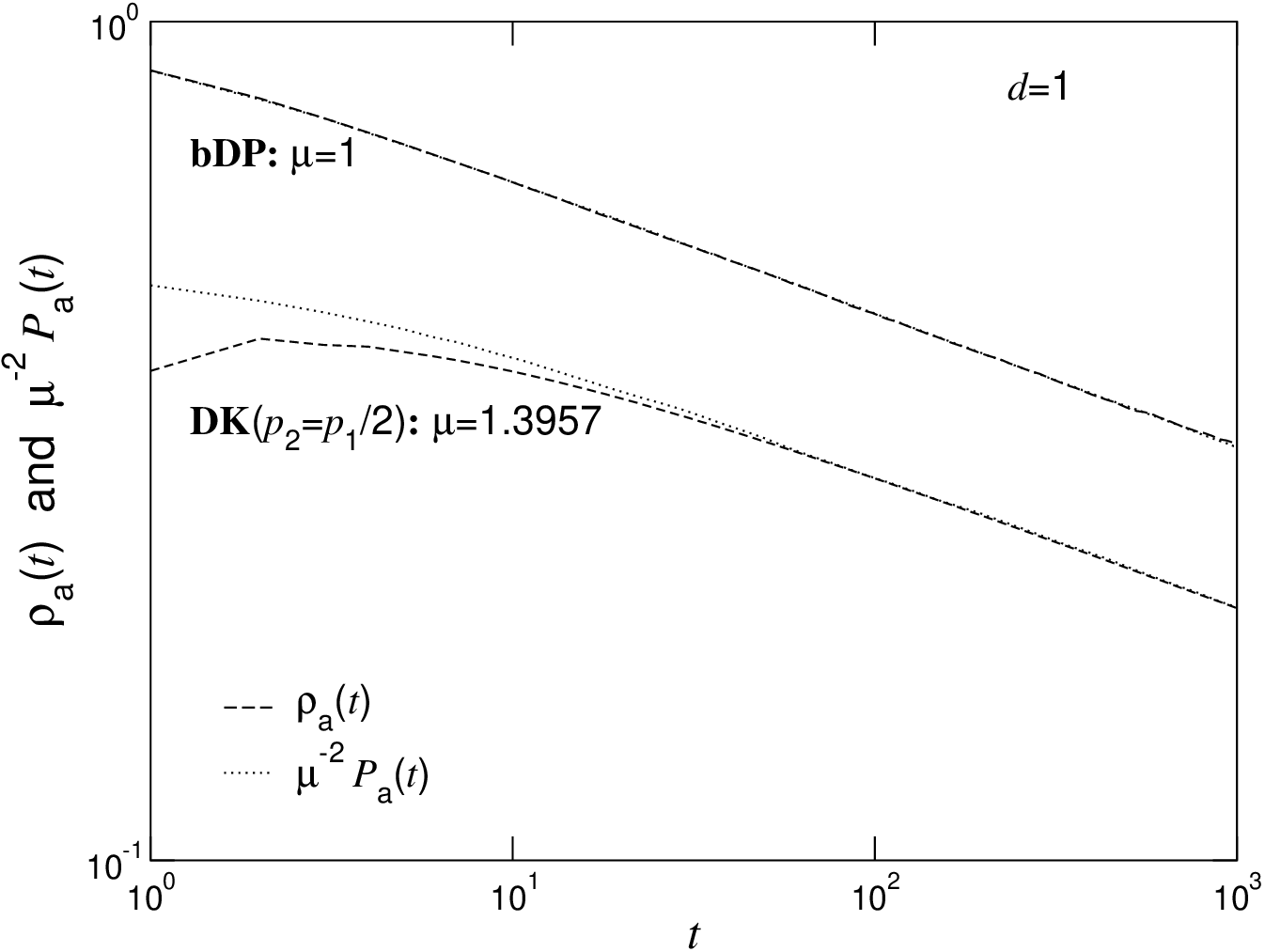}
  \caption{
     The order parameter decay~$\rho_{\scriptscriptstyle{\mathrm{a}}}(t)$ 
     (started from a fully occupied lattice) and the survival 
     probability~$P_{\scriptscriptstyle\mathrm{a}}(t)$
     (started from a single seed) for two distinct critical points
     of the Domany-Kinzel automaton.
     For bond-DP the dual symmetry is fulfilled microscopically 
     leading to $\rho_{\scriptscriptstyle{\mathrm{a}}}(t)
     =P_{\scriptscriptstyle\mathrm{a}}(t)$.
     In general, this equivalence is only asymptotically obeyed
     i.e., $\rho_{\scriptscriptstyle{\mathrm{a}}}(t) \sim 
     \mu^2 P_{\scriptscriptstyle\mathrm{a}}(t)$. 
     Here, the curves for 
     $p_{\scriptscriptstyle 2} = p_{\scriptscriptstyle 1}/2$ are
     shown.
   }
  \label{fig:dual_sym_01} 
\end{figure}

As already mentioned above, all critical points along the
phase transition line (except of the termination point) belong 
to the universality class of directed percolation.
Thus the scaling behavior at all determined critical points
has to obey the rapidity reversal and 
equation\,(\ref{eq:p_sur_alpha_rho}) has to be fulfilled.
In order to check its validity we have measured first the 
order parameter decay 
(started from a fully occupied lattice)
and second the survival probability (of an initially 
single seed) at the critical point.
Examples are shown in figure\,\ref{fig:dual_sym_01}
for $p_{\scriptscriptstyle 2}=p_{\scriptscriptstyle 1}
(2-p_{\scriptscriptstyle 1})$ (bond directed percolation)
as well as for $p_{\scriptscriptstyle 2}=p_{\scriptscriptstyle 1}/2$.
Bond directed percolation is a particular case since the 
rapidity reversal is microscopically obeyed (see e.g.~\cite{HINRICHSEN_1}).
Therefore, equation\,(\ref{eq:p_sur_alpha_rho}) holds for all~$t$ with $\mu=1$, i.e.,
\begin{equation}
P_{\scriptscriptstyle{\mathrm{a}}}(t) \, 
=
\, \rho_{\scriptscriptstyle\mathrm{a}}(t) \, .
\label{eq:p_sur_alpha_rho_bdp}
\end{equation}
This can be seen in figure\,\ref{fig:dual_sym_01} where the survival
probability $P_{\scriptscriptstyle{\mathrm{a}}}(t)$ and the 
order parameter $\rho_{\scriptscriptstyle\mathrm{a}}(t)$ collapse onto 
the same function.

\begin{figure}[t]
  \centering
  \includegraphics[clip,width=7.0cm,angle=0]{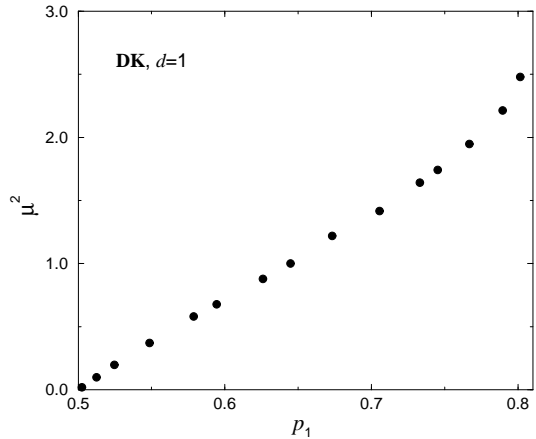}
  \caption{
     Moving along the critical line of the phase diagram,
     the parameter~$\mu$ decreases
     (see right figure) until it vanishes at the termination point,
     reflecting the violation of the rapidity reversal for 
     compact directed percolation.
     Thus~$\mu$ parameterizes the critical line of the one-dimensional
     DK automaton.
   }
  \label{fig:dual_sym_02} 
\end{figure}

Except of this particular case, the order parameter decay equals
the survival probability behavior only asymptotically.
Furthermore, the variable $\mu$ is non-trivial, i.e., $\mu \neq 1$.
Both effects can be seen in figure\,\ref{fig:dual_sym_01}
for $p_{\scriptscriptstyle 2} = p_{\scriptscriptstyle 1}/2$.
After a certain transient $P_{\scriptscriptstyle{\mathrm{a}}}(t)$ and
$\rho_{\scriptscriptstyle{\mathrm{a}}}(t)$ are characterized by the 
same power-law decay, i.e., $\alpha=\delta$. 
Choosing the correct value of $\mu$ (here $\mu\approx 1.4$) 
the tails of both functions collapse onto the same curve.
Worth mentioning, the transient regime increases by approaching the 
termination point.
But more important, the factor $\mu$ varies along the transition
line. 
Moving along the critical line into the direction of the 
CDP termination point, $\mu$ decreases.
Approaching the termination point, $\mu$ tends to zero signalling
the violation of equation\,(\ref{eq:p_sur_alpha_rho}).
In other words, $\mu \to 0$ reflects the breakdown of the 
rapidity reversal caused by the 
change of the universality class from DP to CDP.
In figure\,\ref{fig:dual_sym_02}, $\mu^2$ is plotted as a function
of the distance from the termination point along the transition line.
Notice that $\mu^2$ vanishes linearly as a function 
of $p_{\scriptscriptstyle 1}$. 
In that way, $\mu$ can be used to parameterize the critical line of the 
one-dimensional DK automaton alternatively to the crossover 
parameter~$\kappa$.
But in the below presented scaling analysis,
we prefer to use~$\kappa$ since the
accuracy of the determination of~$\kappa$, i.e., of 
the critical line, is significantly higher than that of~$\mu$.

\begin{table}[ht]
\caption{Estimates of critical points of the $1+1$-dimensional
Domany-Kinzel automaton. Bond directed percolation 
($p_{\scriptscriptstyle 2}=p_{\scriptscriptstyle 1}
(2-p_{\scriptscriptstyle 1})$) is denoted by bDP.
Site directed percolation corresponds to $k=1$.
The data of \protect\cite{JENSEN_16} are obtained by series expansions. 
In $1+1$-dimension, this method yields data of significantly higher 
accuracy than usual Monte Carlo methods.
}
\label{table:pc_werte}
\begin{indented}
\item[]
\begin{tabular}{llll}
\br
$k$  & $p_{\scriptscriptstyle 1,\mathrm{c}}$  &
$p_{\scriptscriptstyle 2,\mathrm{c}}= k 
p_{\scriptscriptstyle 1,\mathrm{c}} $ & $\mu$ \cr  
\mr
0 	& $0.811(1)$  		& $0$ 		&       \cr
0.1 	& $0.8015(4)$ 		& $0.08015$ 	& 1.5747(34)\cr
0.25	& $0.7894(3)$ 		& $0.19735$ 	& 1.4876(32)\cr
0.5	& $0.7668(2)$ 		& $0.3834$ 	& 1.3957(31)\cr
0.7	& $0.74515(7)$ 		& $0.521605$ 	& 1.3195(30)\cr
0.8	& $0.73300(10)$		& $0.5864$ 	& 1.2809(34)\cr
1.0	& $0.70548515(20)$\protect\cite{JENSEN_16}  	
& $0.70548515$ 		& 1.1902(30)\cr
1.2	& $0.67316(11)$		& $0.807792$ 	& 1.1046(26)\cr
bDP 	& $0.644700185(5)$\,\protect\cite{JENSEN_16}
 	& $0.873762052$ 	& 1\cr
1.45	& $0.62585(9)$		& $0.9074825$ 	& 0.93737(20)\cr
1.6	& $0.594305(15)$	& $0.950888$ 	& 0.82363(22)\cr
1.67	& $0.57870(8)$ 		& $0.966429$ 	& 0.76165(26)\cr
1.8	& $0.54865(7)$ 		& $0.98757$ 	& 0.60868(24)\cr
1.9	& $0.52469(6)$ 		& $0.996911$ 	& 0.44501(30)\cr
1.95	& $0.5124250(15)$	& $0.99922875$ 	& 0.31626(38)\cr
1.99	& $0.5024969(15)$ 	& $0.99996903$ 	& 0.14060(44)\cr
\br
\end{tabular}
\end{indented}
\end{table}

Now we consider the crossover scaling behavior using dynamical
simulations.
Various quantities, for example 
$\rho_{\scriptscriptstyle\mathrm{a}}(t)$ as well as 
$P_{\scriptscriptstyle\mathrm{a}}(t)$, reflect the crossover
and are therefore well suited for our analysis.
But in contrast to the survival probability 
($\delta_{\scriptscriptstyle \mathrm{CDP}}> 0$)
the order parameter decay $\rho_{\scriptscriptstyle\mathrm{a}}(t)$
has the advantage
that the rescaling affects only one axis since
$\beta_{\scriptscriptstyle \mathrm{CDP}}=
\alpha_{\scriptscriptstyle \mathrm{CDP}}=0$.
Thus, for the sake of simplicity, we focus our attention to 
the scaling behavior of the order parameter decay for critical 
points along the transition line.
In order to describe the temporal evolution
we incorporate an additional scaling field to the 
steady state scaling form (\ref{eq:scal_cross_01})
\begin{equation}
\rho_{\scriptscriptstyle\mathrm{a}}(\tau,t,\kappa) 
\, \sim \, \lambda^{-\beta_{\scriptscriptstyle \mathrm{CDP}}}
\, {\tilde r}(\tau \, \lambda, 
t \lambda^{-\nu_{\scriptscriptstyle \parallel,\mathrm{CDP}}},
\kappa \, \lambda^{\phi}) \, .
\label{eq:scal_cross_dyn_01}
\end{equation}
Along the transition line ($\tau=0$), the order parameter has to
obey the scaling form 
\begin{equation}
\rho_{\scriptscriptstyle\mathrm{a}}(0,t,\kappa) 
\, \sim \, {\tilde r}(0, 1,
\kappa \, t^{\phi/2})
\label{eq:scal_cross_dyn_02}
\end{equation}
where we made used of the analytical results 
$\beta_{\scriptscriptstyle \mathrm{CDP}}=0$,
$\nu_{\scriptscriptstyle \parallel,\mathrm{CDP}}=2$
and where we have set 
$t\lambda^{-\nu_{\scriptscriptstyle \parallel,\mathrm{CDP}}}=1$.

\begin{figure}
  \centering
  \includegraphics[clip,width=7.0cm,angle=0]{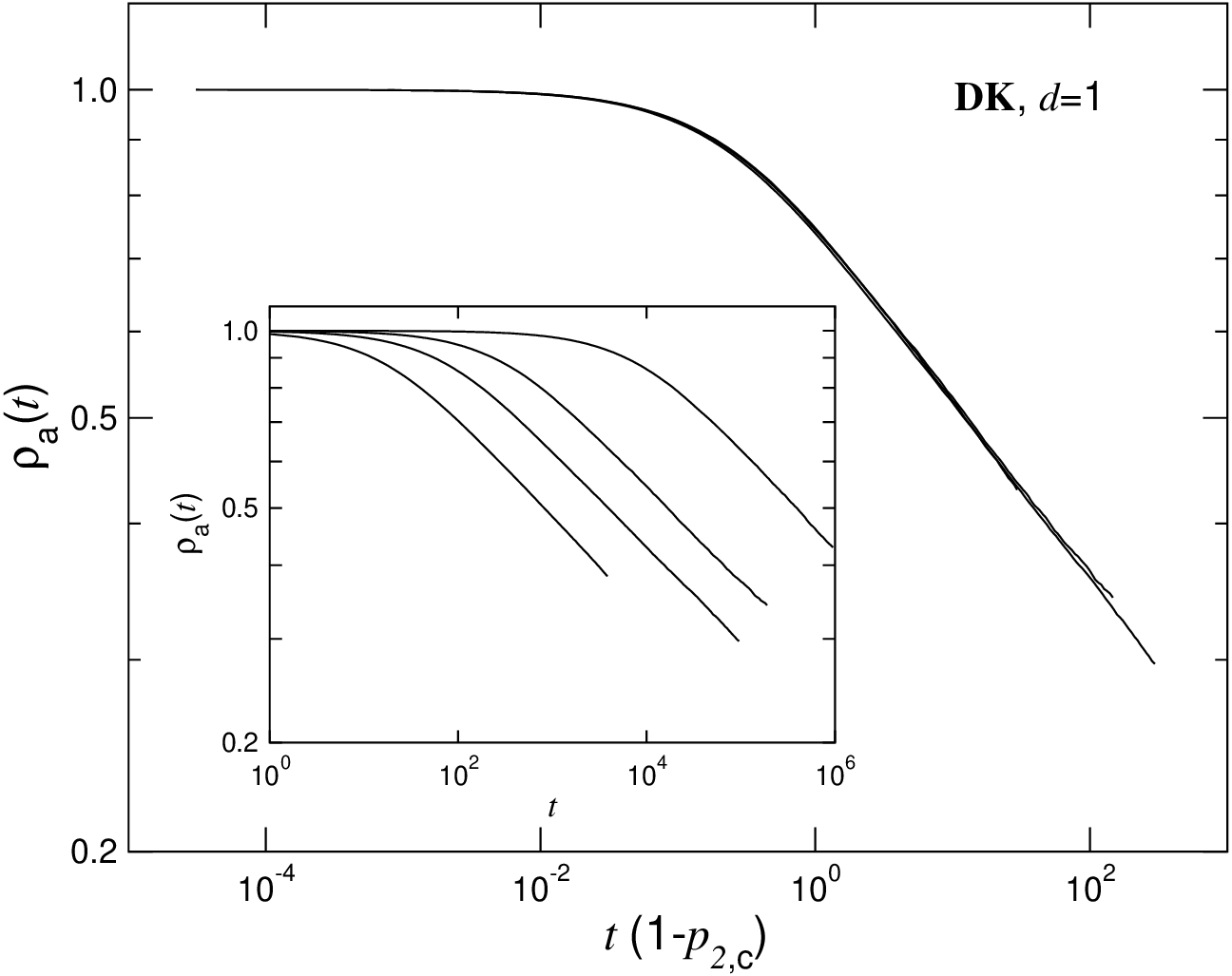}
  \caption{
     The order parameter decay along the critical line of the
     one-dimensional Domany-Kinzel (DK) automaton.
     Unscaled data is shown in the inset
     (for $k=1.8, 1.9, 1.95, 1.99$, from left to right).
     A data collapse is obtained by 
     plotting the order parameter~$\rho_{\scriptscriptstyle \mathrm{a}}$ 
     as a function of the rescaled time 
     $t \kappa \propto t (1-p_{\scriptscriptstyle 2})$.
     Both asymptotic scaling regimes 
     ($\alpha_{\scriptscriptstyle \mathrm{CDP}}=0$ 
     and $\alpha_{\scriptscriptstyle \mathrm{DP}}\approx 0.159$)
     are recovered.
   }
  \label{fig:cross_dp_cdp} 
\end{figure}

The inset of figure\,\ref{fig:cross_dp_cdp} displays the order 
parameter decay close to the termination point 
($k=1.8, 1.9, 1.95, 1.99$ see Table\,\ref{table:pc_werte}).
Starting from the fully occupied lattice the order parameter
remains nearly constant ($\alpha_{\scriptscriptstyle \mathrm{CDP}}=0$) 
before a crossover to the directed
percolation power-law behavior takes place
($\alpha_{\scriptscriptstyle \mathrm{DP}}\approx 0.159$).
The first transient regime increases by approaching the 
termination point.
Using the field theoretical result $\phi=2$~\cite{JANSSEN_4}
these different curves have to collapse onto a single curve
by plotting $\rho_{\scriptscriptstyle\mathrm{a}}$ as a function
of $\kappa t$.
The obtained data collapse (see figure\,\ref{fig:cross_dp_cdp})
confirms the field theoretical result.

\section{Summary}
\label{sec:sum}

We have investigated the crossover scaling behavior 
from directed percolation to compact directed percolation
within the Domany-Kinzel automaton.
Our results confirm the field theoretical results, in
particular the crossover exponent $\phi=2/d$.
Furthermore, we have analysed the rapidity reversal
symmetry along the transition line.
To our knowledge, the important field theoretical 
parameter~$\mu$ is numerically determined for the
first time. 
Approaching the termination point this parameter vanishes, 
reflecting the breakdown of the rapidity reversal.

For the sake of completeness we mention that the crossover from
DP to CDP along the transition line is also reflected in the 
morphology of the spreading clusters. 
As pointed out in~\cite{HINRICHSEN_1}, the varying morphology
corresponds to the change of short-range correlations, leaving 
the long-range correlations (i.e., the scaling behavior) unchanged
unless the termination point is reached.

I would like to thank 
R.\,D.~Willmann for usefull discussions.


\section*{References}

\begin{thebibliography}{10}

\bibitem{MARRO_1}
J. Marro and R. Dickman, {\em Nonequilibrium phase transitions in lattice
  models} (Cambridge University Press, Cambridge, 1999).

\bibitem{HINRICHSEN_1}
H. Hinrichsen, Adv.~Phys. {\bf 49},  815  (2000).

\bibitem{ODOR_1}
G. {\'O}dor, Rev.~Mod.~Phys. {\bf 76},  663  (2004).

\bibitem{LUEB_35}
S. L{\protect\"u}beck, Int.~J.~Mod.~Phys.~B {\bf 18},  3977  (2004).

\bibitem{MENDES_2}
{J.\,F.\,F.~Mendes}, R. Dickman, and H. Herrmann, Phys.~Rev.~E {\bf 54},  3071
  (1996).

\bibitem{DOMANY_1}
E. Domany and W. Kinzel, Phys.~Rev.~Lett. {\bf 53},  311  (1984).

\bibitem{JANSSEN_4}
{H.-K.~Janssen}, J.~Phys.:~Cond.~Mat. {\bf 17},  S1973  (2005).

\bibitem{JANSSEN_1}
{H.-K.~Janssen}, Z.~Phys.~B {\bf 42},  151  (1981).

\bibitem{KAMPEN_1}
{N.\,G.~van\,Kampen}, {\em Stochastic processes in physics and chemistry}
  (North Holland, Amsterdam, 1992).

\bibitem{OHTSUKI_1}
T. Ohtsuki and T. Keyes, Phys.~Rev.~A {\bf 35},  2697  (1987).

\bibitem{OHTSUKI_2}
T. Ohtsuki and T. Keyes, Phys.~Rev.~A {\bf 36},  4434  (1987).

\bibitem{LUEB_31}
S. L{\protect\"u}beck, J.~Stat.~Phys. {\bf 123},  193  (2006).

\bibitem{GRASSBERGER_13}
P. Grassberger, J.~Stat.~Mech. {\bf ~},  P01004  (2006).

\bibitem{CHAICHIAN_1}
M. Chaichian and A. Demichev, {\em Path integrals in physics,
  {$\mathrm{Volume}\;1$}} (Institute of Physics, Bristol, 2001).

\bibitem{JANSSEN_7}
{H.-K.~Janssen}, J.~Stat.~Phys. {\bf 103},  801  (2001).

\bibitem{DEDOMINICIS_1}
{C.\,J.~De\,Dominicis}, J.~Phys.~C~(France) {\bf 37},  247  (1976).

\bibitem{JANSSEN_8}
{H.-K.~Janssen}, Z.~Phys.~B {\bf 23},  377  (1976).

\bibitem{MARTIN_1}
{P.\,C.~Martin}, {E.\,D.~Siggia}, and {H.\,A.~Rose}, Phys.~Rev.~A {\bf 8},  423
   (1973).

\bibitem{LIGGETT_1}
{T.\,M.~Liggett}, {\em Interacting particle systems} (Springer, New York,
  1985).

\bibitem{GRASSBERGER_4}
P. Grassberger and {A.~de\,la\,Torre}, Ann.~Phys.~(N.Y.) {\bf 122},  373
  (1979).

\bibitem{GRASSBERGER_2}
P. Grassberger, Z.~Phys.~B {\bf 47},  365  (1982).

\bibitem{HARRIS_2}
{T.\,E.~Harris}, Ann.~Prob. {\bf 2},  969  (1974).

\bibitem{JENSEN_2}
I. Jensen, Phys.~Rev.~Lett. {\bf 70},  1465  (1993).

\bibitem{ZIFF_1}
{R.\,M.~Ziff}, E. Gulari, and Y. Barshad, Phys.~Rev.~Lett. {\bf 56},  2553
  (1986).

\bibitem{MENDES_1}
{J.\,F.\,F.~Mendes}, R. Dickman, M. Henkel, and {M.\,C.~Marques}, J.~Phys.~A
  {\bf 27},  3019  (1994).

\bibitem{LUEB_32}
S. L{\protect\"u}beck and {R.\,D.~Willmann}, Nucl.~Phys.~B {\bf 718},  341
  (2005).

\bibitem{WOLFRAM_1}
S. Wolfram, Rev.~Mod.~Phys. {\bf 55},  601  (1983).

\bibitem{ZEBENDE_1}
{G.\,F.~Zebende} and {T.\,J.\,P.~Penna}, J.~Stat.~Phys. {\bf 74},  1273
  (1994).

\bibitem{KINZEL_2}
W. Kinzel, Z.~Phys.~B {\bf 58},  229  (1985).

\bibitem{TOME_1}
T. Tom{\'e}, Physica~A {\bf 212},  99  (1994).

\bibitem{RIEGER_1}
H. Rieger, A. Schadschneider, and M. Schreckenberg, J.~Phys.~A {\bf 27},  L423
  (1994).

\bibitem{ATMAN_1}
{A.\,P.\,F.~Atman}, R. Dickman, and {J.\,G.~Moreira}, Phys.~Rev.~E {\bf 67},
  016107  (2003).

\bibitem{ESSAM_1}
{J.\,W.~Essam}, J.~Phys.~A {\bf 22},  4927  (1989).

\bibitem{KEARNEY_1}
{M.\,J.~Kearney} and {R.\,J.~Martin}, J.~Phys.~A: Math.~Gen. {\bf 36},  6629
  (2003).

\bibitem{JENSEN_5}
I. Jensen, J.~Phys.~A {\bf 32},  5233  (1999).

\bibitem{RIEDEL_2}
{E.\,K.~Riedel} and {F.\,J.~Wegner}, Z.~Phys. {\bf 225},  195  (1969).

\bibitem{JENSEN_16}
I. Jensen, J.~Phys.~A {\bf 37},  6899  (2004).

\end{thebibliography}

\end{document}